\documentclass[preprint,nofootinbib,floatfix,a4paper,prd,superscriptaddress]{revtex4-1}   	
\usepackage{geometry}                		
\usepackage[colorlinks,citecolor=blue]{hyperref}
\usepackage{amsmath}
\usepackage{mathrsfs}
\usepackage{bbm}
\usepackage{amsfonts}
\usepackage{amssymb}
\usepackage{latexsym}
\usepackage{graphicx}
\usepackage[english]{babel}
\usepackage{multirow}
\usepackage{float}
\usepackage{url}
\usepackage{slashed}
\usepackage{ulem}

\usepackage{orcidlink}

\newcommand{\be}{\begin{equation}}
\newcommand{\ee}{\end{equation}}
\newcommand{\ba}{\begin{array}}
\newcommand{\ea}{\end{array}}
\newcommand{\bea}{\begin{eqnarray}}
\newcommand{\eea}{\end{eqnarray}}

\usepackage[utf8]{inputenc}
\usepackage[T1]{fontenc}
\usepackage{CormorantGaramond}

\begin{document}

\title{Exploring Abelian--Non-Abelian Kinetic Mixing in SMEFT and Beyond}
\author{Van Que Tran\orcidlink{0000-0003-4643-4050}}
\email{vqtran@gate.sinica.edu.tw}
\affiliation{\small Institute of Physics, Academia Sinica, Nangang, Taipei 11529, Taiwan}
\affiliation{\small Phenikaa Institute for Advanced Study, Phenikaa University, Yen Nghia, Ha Dong, Hanoi 100000, Vietnam}

\author{Tzu-Chiang Yuan\orcidlink{0000-0001-8546-5031} }
\email{tcyuan@phys.sinica.edu.tw}
\affiliation{\small Institute of Physics, Academia Sinica, Nangang, Taipei 11529, Taiwan}

\begin{abstract}

We explore a novel scenario involving Abelian--non-Abelian kinetic mixing within the framework of the Standard Model Effective Field Theory (SMEFT) and its extension with a real triplet scalar field. In SMEFT, this mixing arises exclusively from a dimension-6 operator involving the Standard Model Higgs doublet, while the real triplet scalar field introduces an additional dimension-5 operator. We derive the modifications to electroweak gauge boson properties and impose constraints using electroweak precision data.
In SMEFT, we find that $Z$ pole data at LEP-I imposes a stringent constraint on the kinetic mixing parameter, requiring it to be less than $\cal O$($10^{-4}$), which corresponds to a new physics scale of about 10 TeV. 
In the SMEFT+triplet scenario, the constraint can be significantly relaxed with a sizeable triplet vacuum expectation value while preserving custodial symmetry in a finely-tuned parameter space.
Future measurements from the Circular Electron Positron Collider could probe the kinetic mixing parameter down to an order of magnitude smaller.

\end{abstract}
\maketitle

\section{Introduction}
\label{sec:intro}

Kinetic mixing between different gauge groups is a well-established concept in extensions of the Standard Model (SM). The most widely studied scenario involves kinetic mixing between the SM hypercharge and a new $U(1)$ gauge group~\cite{Holdom:1985ag}. This interaction enables a dark photon, the gauge boson associated with the new $U(1)$, to interact with visible particles in the SM, effectively bridging the dark sector and the SM sector. Such mixing is typically described by renormalizable operators and has been extensively explored in the literature~\cite{Holdom:1986eq,Feldman:2007wj,Curtin:2018mvb,Feng:2022inv}.

Beyond this Abelian kinetic mixing, a possibility involves kinetic mixing between Abelian and non-Abelian gauge fields, specifically between the $U(1)$ and $SU(2)$ gauge groups, hereafter referred to as Abelian--non-Abelian kinetic mixing.
Unlike the Abelian case, this type of mixing arises from non-renormalizable operators, which are naturally described in the Effective Field Theory (EFT) framework.  
In the context of the Standard Model Effective Field Theory (SMEFT)~\cite{Weinberg:1979sa,Buchmuller:1985jz,Grzadkowski:2010es,Brivio:2017vri}, such kinetic mixing can be induced by a dimension-6 operator involving the SM Higgs doublet $\Phi$. The presence of this kinetic mixing leads to modifications in the properties of the electroweak gauge bosons, which can be constrained by precision electroweak measurements~\cite{GRINSTEIN1991326, Han:2004az,MANOHAR2006107}.
Moreover, 
previous studies have predominantly focused on models involving an extended dark gauge sector, where kinetic mixing occurs either between a dark $U(1)$ and the SM $SU(2)_L$~\cite{Barello:2015bhq, Arguelles:2016ney}, or between the SM $U(1)$ hypercharge and a dark $SU(2)$~\cite{Zhang:2009dd, Chen:2009ab, Ko:2020qlt, Nomura:2021tmi, Zhou:2022pom}.

In this work, we investigate Abelian--non-Abelian kinetic mixing within the SM gauge groups themselves, 
specifically focusing on the mixing between hypercharge $U(1)_Y$ and $SU(2)_L$ and its implications for electroweak precision observables.
In addition to the dimension-6 operator, we consider a dimension-5 operator involving an extension to the SMEFT with a real triplet scalar field $\Sigma$.

In the next section, we formulate the notion of Abelian--non-Abelian kinetic mixing in SMEFT and its simple extension with a real triplet scalar. Constraints from precision electroweak data is presented in Section~\ref{sec:constraints}. We summarize our findings in Section~\ref{sec:summary}.

\section{Abelian--non-Abelian Kinetic Mixing}
\label{sec:kmix}

 \subsection{SMEFT dimension-6 operator}
 Within the SMEFT, the Abelian--non-Abelian kinetic mixing can be generated by dimension-6 operator 
 \be
 \label{eq:O6}
 {\cal O}_6 = \frac{ g g^\prime c_6  }{\Lambda_6^2}  \mathrm{Tr} \left[ \Phi^{\dagger} W_{\mu \nu} \Phi \right]   B^{\mu \nu},
 \ee
 where $g$ ($g^{\prime}$) is the gauge coupling of $SU(2)_L$ ($U(1)_Y$), $c_6$ is dimensionless coefficient, $\Lambda_6$ is the new physics scale, $\Phi$ is the SM Higgs doublet, and $W_{\mu \nu} = W_{\mu\nu}^a \frac{ \tau^a }{2}$ and $B^{\mu \nu}$ are the $SU(2)_L$ and $U(1)_Y$ field strength tensors, respectively. 
 Here $\tau^a \, (a=1,2,3)$ are the Pauli matrices. The trace $\mathrm{Tr}$ is taking over the $SU(2)_L$ space.
 We note that ${\cal O}_6$ belongs to the class of dimension-6 operators in SMEFT~\cite{Buchmuller:1985jz,Grzadkowski:2010es}.

 After electroweak spontaneously symmetry breaking, the doublet scalar can take the form of 
 \be
 \label{eq:scalarfields}
 \Phi = 
 \begin{pmatrix}
 G^+ \\ \frac{1}{\sqrt 2} \left( v_\Phi + \phi + i G^0 \right)
 \end{pmatrix}, 
 \ee
 where $G^\pm$ and $G^0$ are Goldstone bosons, $\phi$ is the physical Higgs boson,   
 and $v_\Phi$ is the vacuum expectation value (VEV) of the $SU(2)_L$ doublet scalar.
 The operator in Eq.~(\ref{eq:O6}) induces the following mixing term
 \be
 \label{eq:kinmixing}
 - \frac{1}{2} \epsilon W_{\mu\nu}^3 B^{\mu\nu} 
 \ee
 where the kinetic mixing parameter $\epsilon \equiv \epsilon_6 = g g^\prime c_6  v_\Phi^2/(2 \Lambda_6^2)$.  
 
 The kinetic energy Lagrangian for the gauge sector can then be given as
 \be
 \label{eq:kin}
 {\cal L}_{\rm kin} =  -\frac{1}{4} {B}_{\mu\nu} {B}^{\mu\nu} - \frac{1}{4} {W}^a_{\mu\nu} {W}^{\mu\nu a} 
 - \frac{1}{2} \epsilon {W}^3_{\mu\nu} {B}^{\mu\nu} .
 \ee
 To eliminate the kinetic mixing term in Eq.~(\ref{eq:kin}), one can apply the following field redefinition
\be
\label{eq:fieldredefinition}
\begin{pmatrix}
    {W}^3_\mu \\
    {B}_\mu 
    \end{pmatrix}
    \to  
    \begin{pmatrix}
    \frac{1}{\sqrt{1-\epsilon^2}} W^3_{\mu} \\
    B_\mu - \frac{\epsilon}{\sqrt{1-\epsilon^2}} W^3_{\mu}
\end{pmatrix}.
\ee
This redefinition affects the covariant derivative, which now takes the form
\be
\label{eq:covariantD}
D_{\mu}  = \partial_\mu - i g ( T^1 W^{1}_\mu +  T^2 W^{2}_\mu ) - i \left(\frac{g T^3 - \epsilon g' Y}{\sqrt{1-\epsilon^2}} \right) W^3_\mu - i g^{\prime} Y B_{\mu} 
\ee
where $T^a$ with $a=1,2,3$ are the $SU(2)_L$ generators, $Y$ is the hypercharge. For doublet, $(T^a)_{ij}= (\tau^a)_{ij}/2$ with $i,j=1,2$; while for triplet $(T^a)_{bc} = - i \epsilon_{abc}$ with $a,b,c=1,2,3$.

The Abelian--non-Abelian kinetic mixing affects only the third term in Eq.~(\ref{eq:covariantD}), leading to a modification in the $Z$ boson mass
\be
m_Z^2 = \frac{1}{4} (\tilde{g}^2 + g'^2) v_\Phi^2,
\ee
where 
$\tilde{g} = (g + \epsilon g')/\sqrt{1-\epsilon^2}$,  
while 
the charged $W^\pm$ gauge boson mass 
$m_{W^\pm} = \frac{1}{2} g v_\Phi$ stays unchanged and 
the photon field $A$ remains massless. The weak mixing angle is now given by
\be
\tilde{c}_w = \frac{\tilde g} {\sqrt{\tilde{g}^2 +g'^2}} ,
\ee
where $\tilde{c}_w$ ($\tilde{s}_w$) $\equiv \cos {\tilde{\theta}_w}$ ($\sin {\tilde{\theta}_w}$).
Hence $Z_\mu = - \tilde s_w B_\mu + \tilde c_w W^3_\mu$ and $A_\mu = \tilde c_w B_\mu + \tilde s_w W^3_\mu$.

The covariant derivative in Eq.~(\ref{eq:covariantD}) can then be rewritten in terms of physical fields 
\be
\label{eq:covariantD2}
D_{\mu}  = \partial_\mu - i \frac{g}{\sqrt{2}} (W_\mu^+ T^+  + W_\mu^- T^-)  
		- i  g_M Z_\mu \left[T^3 - \tilde{s}_w \left( \tilde{s}_w  + \frac{\epsilon}{ \sqrt{1-\epsilon^2} } \tilde{c}_w \right)Q \right]  
		- i e A_{\mu} Q ,
\ee
where $T^\pm = T^1 \pm i T^2$, $g_M = \sqrt{\tilde{g}^2 + g'^2}$, $Q = T^3 + Y$ and the electrical charge $e$ is given by~\footnote{
Alternatively, one can define   
\be
\frac{1}{e^2} = \frac{1}{g'^2} + \frac{1 + 2 \epsilon g/g'}{g^2}, \nonumber
\ee
and then identifying $g'_{\rm SM} \equiv g'$ and $g_{\rm SM} \equiv \frac{g}{\sqrt{1 + 2\epsilon g/g'}}$.
}
\be
\frac{1}{e^2} = \frac{1}{g^2} + \frac{1 + 2 \epsilon g'/g}{g'^2}.
\ee
We thus identify 
\be
g_{\rm SM} \equiv g \;\;\; {\rm and} \;\;\; g'_{\rm SM} \equiv \frac{g'}{\sqrt{1 + 2\epsilon g'/g}}. 
\ee

The neutral current interaction Lagrangian in the model, involving the couplings of SM fermions to the photon and $Z$ boson fields, is given by
\be
\label{eq:LN}
{\cal L}_N = g_M j^\mu_Z Z_\mu + e  j^\mu_{\rm EM} A_\mu ,
\ee
where $j^\mu_Z = \sum_f \bar{f} \gamma^\mu (v_f - \gamma_5 a_f) f $  with 
\be
\label{eq:vfaf}
v_f =  \left [ \frac{1}{2} T^3_f -  \tilde{s}_w \left( \tilde{s}_w  + \frac{\epsilon}{ \sqrt{1-\epsilon^2} } \tilde{c}_w \right) Q_f \right]  \; \;{\rm and} \;\;
a_f =   \frac{1}{2} T_f^3 ,
\ee
and $j^\mu_{\rm EM} = \sum_f  Q_f \bar{f}\gamma^\mu f $.

\subsection{The dimension-5 operator}
If a real \(SU(2)_L\) triplet scalar is introduced to the SMEFT, it can generate Abelian--non-Abelian kinetic mixing through the dimension-5 operator
 \be
 \label{eq:O5}
 {\cal O}_5 = \frac{ g g^\prime c_5 }{\Lambda_5} \mathrm{Tr} \left[ \Sigma W_{\mu \nu} \right]  B^{\mu \nu} , 
 \ee
 where $c_5$ is a dimensionless coefficient, $\Lambda_5$ is the new physics scale, and $\Sigma$ is the real triplet scalar field with hypercharge zero. The triplet scalar field is conventionally expressed as follows \cite{FileviezPerez:2008bj}:
 \be
 \Sigma =  \Sigma^a \frac{\tau^a }{2} = \frac{1}{2}
 \begin{pmatrix}
  v_\Sigma + \Sigma^0  & \sqrt{2} \Sigma^+ \\  
  \sqrt{2} \Sigma^- & - v_\Sigma - \Sigma^0\\  
  \end{pmatrix},
  \ee
where $\Sigma^\pm$ ($\Sigma^0$) are the charged (neutral) triplet scalar bosons, and $v_\Sigma$ is the VEV of the triplet scalar. 
After electroweak spontaneous symmetry breaking, the operator in Eq.~(\ref{eq:O5}) can induce a kinetic mixing identical to that described in Eq.~(\ref{eq:kinmixing}), with the kinetic mixing parameter defined as $\epsilon \equiv \epsilon_5 = -\frac{ g g^\prime c_5  v_\Sigma}{\Lambda_5}$.

With the presence of the real triplet scalar, in addition to the doublet VEV contribution, as in the SM, the $W^\pm = \frac{1}{\sqrt{2}} (W^1_\mu \mp i W^2_\mu)$ boson mass also receives a contribution from the triplet VEV, leading to
\be
m_{W^\pm}^2 = \frac{1}{4} g^2 (v_\Phi^2 + 4 v_{\Sigma}^2).
\ee
Here $v_\Phi$ and $v_{\Sigma}$ are related to the SM VEV as $v_{\rm SM} = \sqrt{v_{\Phi}^2 + 4 v_{\Sigma}^2 } = 246$ GeV. 
It is important to note that the triplet VEV does not contribute to the $Z$ boson mass in this scenario.

\section{Constraints}
\label{sec:constraints}

In this section, we examine the constraints on Abelian–non-Abelian kinetic mixing and the associated parameter space within two scenarios. The first scenario considers only the \texttt{SMEFT}, where kinetic mixing is generated exclusively through a dimension-6 operator. The second scenario extends beyond the SMEFT framework by incorporating a real $SU(2)_L$ triplet scalar (\texttt{SMEFT+$\Sigma$}). In this case, kinetic mixing arises from both dimension-5 and dimension-6 operators, resulting in a total kinetic mixing parameter given by $\epsilon = \epsilon_5 + \epsilon_6$. We assume a common new physics scale for both operators, such that $\Lambda_5 = \Lambda_6 = \Lambda$.

\subsection{Constraint from $\rho$ parameter}
As shown in the previous section, the tree-level masses of the $Z$ and $W$ bosons can deviate from their SM values due to the presence of the Abelian--non-Abelian kinetic mixing and the triplet scalar field. This deviation implies a possible violation of custodial symmetry at tree level, leading to corrections to the well-known $\rho$ parameter, 
\be
\rho = \frac{m_W^2}{m_Z^2 \tilde{c}^2_w} = 1 + \delta\rho,
\ee
where 
\be
\delta\rho \simeq - 2 \epsilon \frac{g_{\rm SM}'}{g_{\rm SM}} 
\ee
in the \texttt{SMEFT} scenario with $\epsilon = \epsilon_6$ and  
\be
\delta\rho \simeq \frac{4v_{\Sigma}^2} {v_\Phi^2} - 2 \epsilon \frac{g_{\rm SM}'}{g_{\rm SM}} \left( 1 + \frac{4v_{\Sigma}^2} {v_\Phi^2} \right)
\ee
in the \texttt{SMEFT+$\Sigma$} scenario with $\epsilon = \epsilon_5 + \epsilon_6$.

The experimental value of $\rho$ is $1.00031 \pm 0.00019$ at $1\sigma$ confidence level~\cite{PhysRevD.110.030001}.
This corresponds to $\delta \rho_{\rm exp} \simeq 0.00031 \pm 0.00019$.
For the \texttt{SMEFT} scenario, this constraint on the kinetic mixing parameter is $\epsilon_6 \lesssim 1.6\times 10^{-4}$, which corresponds to a new physics scale of $\Lambda_6 \gtrsim 6.7$ TeV, assuming the dimensionless coefficient \(c_6 \sim {\cal O}(1)\).

\begin{figure}[tb]
         \centering
	\includegraphics[width=0.55\textwidth]{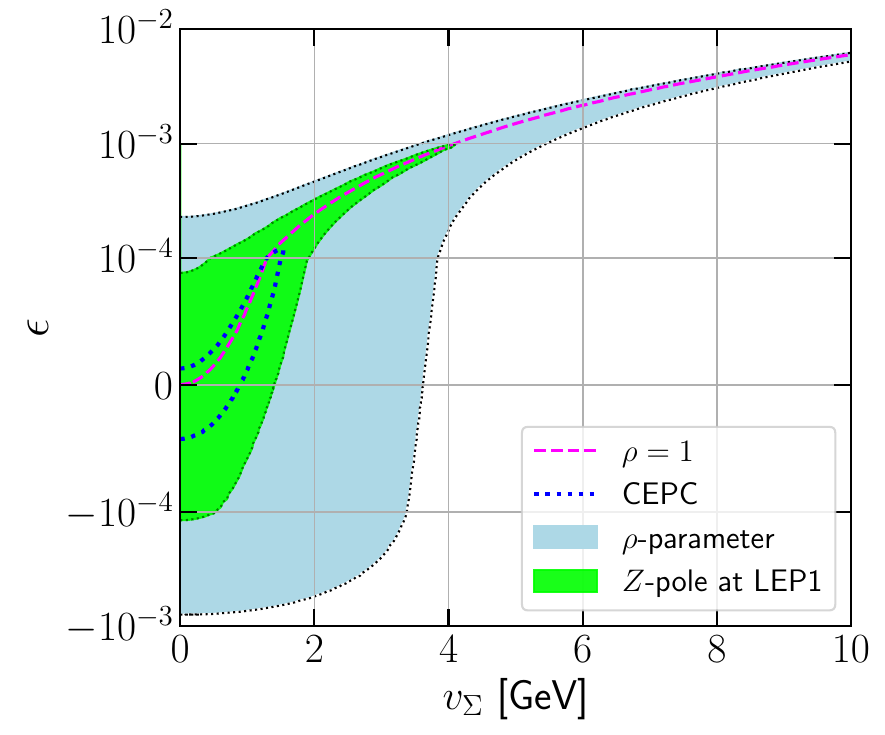}
	\caption{ \label{fig:constraints1} 
	Allowed regions projected onto the ($v_\Sigma$, $\epsilon$) plane in the \texttt{SMEFT+$\Sigma$} scenario. The light blue and green shaded areas correspond to the regions allowed by the $\rho$ parameter measurement \cite{PhysRevD.110.030001} and $Z$ pole physics data \cite{ALEPH:2005ab}, respectively. The dotted blue line indicates the projected region from CEPC electroweak precision data \cite{CEPC-SPPCStudyGroup:2015csa,CEPCStudyGroup:2018ghi}. The magenta line represents $\rho = 1$, where custodial symmetry is maintained.} 
\end{figure}

\begin{figure}[tb]
         \centering
	\includegraphics[width=0.45\textwidth]{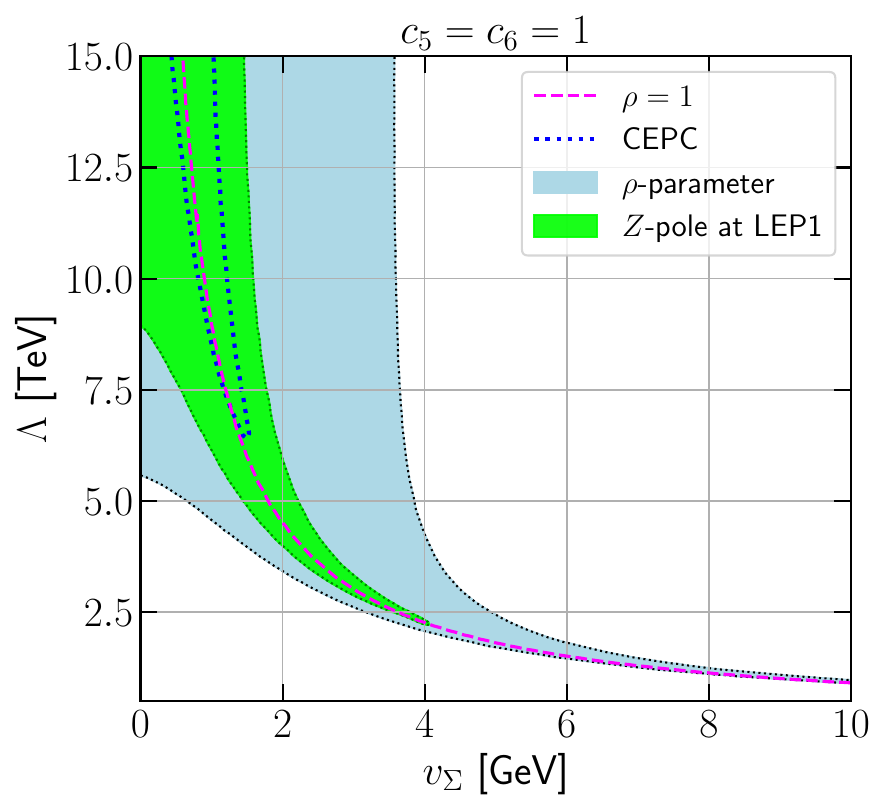}
	\includegraphics[width=0.45\textwidth]{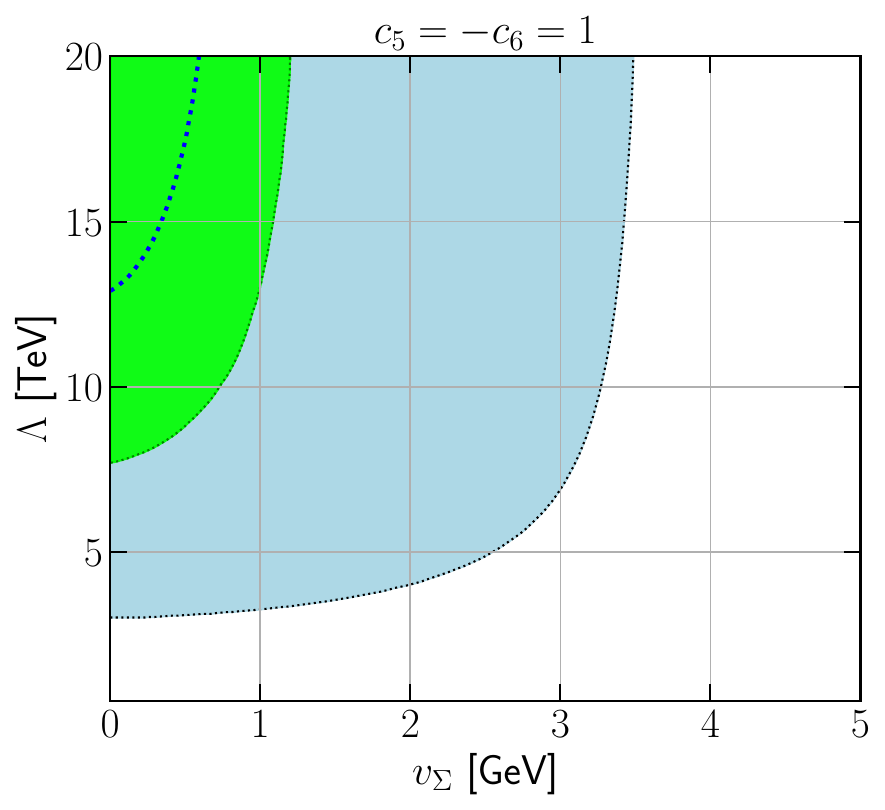}
	\caption{\label{fig:constraints2} Allowed regions projected onto the ($v_\Sigma$, $\Lambda$) plane in the \texttt{SMEFT+$\Sigma$} scenario. The dimensionless coefficients are fixed at $c_5 = c_6 \sim \mathcal{O}(1)$ for the left panel, and $c_5 = -c_6 \sim \mathcal{O}(1)$ for the right panel.}  
\end{figure}

In the \texttt{SMEFT+$\Sigma$} scenario, the regions allowed by the \(\rho\) parameter data are represented by the light blue shaded areas in Fig.~\ref{fig:constraints1}, projected onto the ($v_\Sigma$, $\epsilon$) plane, and in Fig.~\ref{fig:constraints2}, projected onto the ($v_\Sigma$, $\Lambda$) plane. The kinetic mixing parameter $\epsilon$ increases as the triplet VEV $v_\Sigma$ extends to larger values, as shown in Fig.~\ref{fig:constraints1}. For instance, $\epsilon \simeq 0.01$ when $v_\Sigma$ reaches approximately 13 GeV. In the absence of kinetic mixing $(\epsilon \simeq 0)$ or if it has a negative value, $v_\Sigma$ is constrained to $ \lesssim 3.5$ GeV. Interestingly, for $\epsilon \geq 0$ region, custodial symmetry can be preserved at the tree level within this scenario, as demonstrated by the fine-tuned parameter space along the dashed magenta line in Fig.~\ref{fig:constraints1}. 

We observe that an increase in the triplet VEV corresponds to a decrease in the new physics scale $\Lambda$ when the dimensionless coefficients are fixed at $c_5 = c_6 \sim \mathcal{O}(1)$, as shown in the left panel of Fig.~\ref{fig:constraints2}. On the other hand, fixing $c_5 = -c_6 \sim \mathcal{O}(1)$ results in an increase in $\Lambda$ with an increase in the triplet VEV, as illustrated in the right panel of Fig.~\ref{fig:constraints2}. Here, we note that the kinetic mixing parameter is positive when $c_5 = c_6 \sim \mathcal{O}(1)$ and becomes negative when 
$c_5 = -c_6 \sim \mathcal{O}(1)$.

\subsection{Constraints from precision electroweak data at $ Z $ pole physics}
We now explore the constraints on Abelian--non-Abelian kinetic mixing using precision electroweak data at $ Z $ pole physics~\cite{ALEPH:2005ab}. 
The presence of this kinetic mixing modifies the interaction between the $Z$ boson and SM fermions ($Z\bar{f}f $), as detailed in Eq.~(\ref{eq:LN}). 
The couplings of $ Z\bar{f}f $ at tree level, as shown in Eq.~(\ref{eq:vfaf}), can be affected by radiative corrections stemming from propagator self-energies and flavor-specific vertex corrections~\cite{ALEPH:2005ab}. These corrections are incorporated into the couplings as  
\be
\label{eq:vfafbar}
\bar{v}_f =  \sqrt{\rho_f} \left [ \frac{1}{2} T^3_f -  \kappa_f \tilde{s}_w \left( \tilde{s}_w  + \frac{\epsilon}{ \sqrt{1-\epsilon^2} } \tilde{c}_w \right) Q_f \right]  \; \;{\rm and} \;\;
\bar{a}_f =   \frac{1}{2}  \sqrt{\rho_f} T_f^3
\ee
where $\rho_f$ and $\kappa_f$ represent the radiative corrections and are dependent on the fermion \( f \) and the chosen renormalization scheme~\cite{PhysRevD.110.030001}.

In the on-shell renormalization scheme, the decay width of the $Z$ boson into fermion-antifermion pairs is given by~\cite{Baur:2001ze}:
\be
 \Gamma( Z \to f\bar f ) = N_f^c  \mathcal{R}_f   \left( \frac{g_M^2 m_Z}{12 \pi }\right)  \sqrt {1 - 4 \mu _f^2 } 
 \Big [ \vert \bar{v}_f \vert ^2 \left(1 + 2\mu _f^2 \right) + \vert \bar{a}_f \vert^2 \left(1 -4\mu _f^2 \right) \Big] \, , 
 \label{eq:ZwidthRC}
 \ee
where $N_f^c$ is the color factor, with a value of 1 for leptons and 3 for quarks, and $\mu _f = m_f  / m_Z$. 
The factor $\mathcal{R}_f$ in Eq.~(\ref{eq:ZwidthRC}) accounts for QED and QCD corrections and is defined as
\be
 \mathcal{R}_f  = \left(1 + \delta _f^{QED} \right)\left(1 +\frac{N_f^c-1}{2}\delta _f^{QCD} \right) \, ,
 \ee
 where
 \bea
 \delta _{f}^{QED}  &=&\frac{3\alpha}{4\pi}Q_f^2 \, , \\
 \delta _{f}^{QCD}  &=& \frac{{\alpha_s }}{\pi } + 1.409\left( {\frac{{\alpha _s }}{\pi }} \right)^2
 -12.77\left( {\frac{{\alpha _s }}{\pi }} \right)^3  - Q_f^2
\frac{{\alpha \alpha _s   }}{{4\pi ^2 }} \, .
 \eea
Here
$\alpha$ and $\alpha _s $ are the fine-structure and strong coupling constants, 
respectively, evaluated at the $m_Z$ scale.

High precision measurements of other $Z$ pole ($ \sqrt{s} \approx m_Z $) observables are also considered in this analysis, including:

\begin{itemize}
    \item Ratios of the partial decay widths of the \( Z \) boson:
    \begin{equation}
    R_{l} = \frac{\Gamma_{\rm had}}{\Gamma_{l^+l^-}} \, , \hspace{1cm} R_{q} = \frac{\Gamma_{q\bar q}}{\Gamma_{\rm had}} \, ,
    \end{equation}
    where $\Gamma_{\rm had}$ is the partial decay width of $Z$ boson into hadrons and $\Gamma_{l^+l^-}$ ($\Gamma_{q\bar q}$) is the partial decay width of $Z$ boson into a lepton pair $l^+ l^-$ (quark pair $q\bar q$).

    \item Hadronic cross-section:
    \begin{equation}
    \sigma_{\rm had}=  \frac{12\pi \Gamma_{e^+e^-}\Gamma_{\rm had}}{M_Z^2 \, \Gamma_Z^2} \, ,
    \end{equation}

    \item Parity violation quantity:
    \begin{equation}
    A_{f}  =   \frac{2 \bar{v}_f \bar{a}_f}{\bar{v}^2_f + \bar{a}^2_f} \, ,
    \end{equation}

    \item Forward-backward asymmetry quantity:
    \begin{equation}
    A_{\rm FB}  =   \frac{3}{4} A_f \frac{A_e + P_e}{1 + P_e A_e} \, ,
    \end{equation}
where $ P_e $ is the initial $e^-$ polarization. At LEP-I, $P_e = 0$, so the forward-backward asymmetry simplifies to
\begin{equation}
A^{(0,f)}_{\rm FB}  =\frac{3}{4} A_e A_f \, . 
\end{equation}
\end{itemize}
A summary of experimental values of the electroweak observables at the $ Z $ pole from various experiments~\cite{PhysRevD.110.030001} as well as their projected sensitivities at Circular Electron Positron Collider (CEPC)~\cite{CEPCStudyGroup:2018ghi} is provided in Table \ref{table:ewobs}.

\begin{table}[th!]
	\centering		
	\begin{tabular}{ c | c | c | c  }
		\hline
		\hline
		Observable             & Value     & CEPC Uncertainty~\cite{CEPC-SPPCStudyGroup:2015csa, CEPCStudyGroup:2018ghi} & Standard Model \\ 
		\hline
		$M_Z$ [GeV]       & 91.1876 $\pm$ 0.0021   &  $0.0005$  &  91.1884 $\pm$ 0.0019\\
		$\Gamma_Z$ [GeV]       & 2.4955 $\pm$ 0.0023 &  $0.0005$ & 2.4940 $\pm$ 0.0009 \\
		$ \sigma_ {\rm had}[{\rm nb}] $      & 41.481 $\pm$ 0.033  & --- & 41.481 $\pm$ 0.009\\
		$R_e        $        & 20.804 $\pm$ 0.050  &  0.0021 & 20.736$\pm$ 0.010\\
		$R_\mu     $           & 20.784 $\pm$ 0.034  &  $0.0021$  & 20.736 $\pm$ 0.010\\
		$R_\tau   $             & 20.764 $\pm$  0.045 &  $0.0021$ & 20.781 $\pm$ 0.010 \\
		$R_b     $           & 0.21629 $\pm$ 0.00066   &  $4.3\times 10^{-5}$ & 0.21583 $\pm$ 0.00002\\
		$R_c    $            & 0.1721 $\pm$ 0.0030   &  --- & 0.17221 $\pm$ 0.00003\\				
		$A^{(0,e)}_{\rm FB}    $      & 0.0145 $\pm$ 0.0025  &  $8.1\times 10^{-5}$ & 0.01606 $\pm$ 0.00006\\
		$A^{(0,\mu)}_{\rm FB}  $        & 0.0169 $\pm$ 0.0013  & $4.9\times 10^{-5}$ & 0.01606 $\pm$ 0.00006\\
		$A^{(0,\tau)}_{\rm FB}$          & 0.0188 $\pm$ 0.0017   &  --- & 0.01606 $\pm$ 0.00006\\
		$A^{(0,b)}_{\rm FB} $         & 0.0996 $\pm$ 0.0016   &   $0.0001$ & 0.1026 $\pm$ 0.0002\\
		$A^{(0,c)}_{\rm FB} $        & 0.0707 $\pm$ 0.0035    &  --- & 0.0732 $\pm$ 0.0002\\
		$A^{(0,s)}_{\rm FB}$          & 0.0976 $\pm$ 0.0114   &  --- & 0.1027 $\pm$ 0.0002\\
		$A_e   $             & 0.15138 $\pm$ 0.00216  &    --- & 0.1469 $\pm$ 0.0003\\
		$A_\mu $               & 0.142 $\pm$ 0.015   &  --- & 0.1469 $\pm$ 0.0003\\
		$A_\tau$                & 0.136 $\pm$ 0.015  &    --- & 0.1469 $\pm$ 0.0003\\
		$A_b $               & 0.923 $\pm$ 0.020  &   --- & 0.9347\\
		$A_c $               & 0.670 $\pm$ 0.027  &  --- & 0.6674 $\pm$ 0.0001\\
		$A_s$                & 0.0895 $\pm$ 0.091 & --- & 0.9356\\ 
		\hline
		\hline      
	\end{tabular}	
	\caption {Electroweak observables at the $Z$ pole. The second, third, and last columns show the current experimental values \cite{PhysRevD.110.030001}, projected uncertainties from the CEPC preliminary conceptual design reports \cite{CEPC-SPPCStudyGroup:2015csa,CEPCStudyGroup:2018ghi}, and the SM predictions \cite{PhysRevD.110.030001}, respectively.}
	\label{table:ewobs}
\end{table}

We construct a chi-squared function for the electroweak observables at the $Z$ pole as
\be
\label{eq:chisq}
\chi^2 = \sum_i \frac{\left({O}_i^{\rm exp} - O_i^{\rm th}\right)^2} {\left( \delta O_i^{\rm exp} \right)^2}, 
\ee
where $O_i^{\mathrm{exp/th}}$ represents the experimental or theoretical value of the electroweak observables listed in Table~\ref{table:ewobs}, and $\delta O_i^{\mathrm{exp}}$ denotes the corresponding experimental uncertainty. In this analysis, we assume that the theoretical uncertainty is much smaller than the experimental uncertainty, and therefore, we choose to ignore it. 
We then impose the condition $\Delta \chi^2 = \chi^2 - \min(\chi^2) < 5.99$ ($3.84$) to obtain the 2$\sigma$ allowed region on the two-dimensional (one-dimensional) parameter space.

Compared to the constraints from the $\rho$ parameter, these data provide more stringent bounds. For the \texttt{SMEFT} scenario, the current data from high-precision measurements at the $Z$ pole requires the kinetic mixing parameter $\epsilon_6 \lesssim 7\times 10^{-5}$, corresponding to a new physics scale $\Lambda_6 \gtrsim 10$ TeV if $c_6 \sim \mathcal{O}(1)$. Future precision measurements at CEPC can probe the kinetic mixing parameter to about one order of magnitude smaller, thereby extending the new physics scale up to $\sim 32$ TeV.

For the \texttt{SMEFT+$\Sigma$} scenario, the 2$\sigma$ allowed regions on the parameter space from these high-precision data are indicated by the green shaded area in Figs.~\ref{fig:constraints1} and \ref{fig:constraints2}. Specifically, the kinetic mixing is required to be $-10^{-4} \lesssim \epsilon \lesssim 0.001$, and the triplet VEV needs to be $v_\Sigma \lesssim 4$ GeV, as shown in Fig.~\ref{fig:constraints1}. In the limit of $\epsilon \sim 0$, the triplet VEV is further constrained to $v_\Sigma \lesssim 1.5$ GeV. Additionally, these precision measurements from $Z$ pole physics require the new physics scale to be $\Lambda \gtrsim 2$ TeV if one fixes $c_5 = c_6 \sim \mathcal{O}(1)$, as shown in the left panel of Fig.~\ref{fig:constraints2}. 
The bound on $\Lambda$ is more stringent for the case of $c_5 = -c_6 \sim \mathcal{O}(1)$ as shown in the right panel of Fig.~\ref{fig:constraints2}.

Future precision measurements at CEPC will further probe the parameter space, as indicated by the regions bounded by the blue dotted lines in Fig.~\ref{fig:constraints2}. Compared to current $Z$ pole data, these measurements can explore the kinetic mixing parameter to within an order of magnitude smaller. Additionally, they can extend the reach of the new physics scale up to approximately $6.5$ TeV if one assumes $ c_5 = c_6 \sim \mathcal{O}(1)$, and up to around $13$ TeV if $ c_5 = -c_6 \sim \mathcal{O}(1)$.

\section{Conclusions}
\label{sec:summary}

We have studied Abelian--non-Abelian kinetic mixing within and beyond the SMEFT framework. 
In the SMEFT framework, this kinetic mixing can be generated by a dimension-6 operator involving the SM Higgs doublet. 
In contrast, in a scenario involving a real triplet scalar field, this kinetic mixing can arise from a dimension-5 operator. 
We derived the resulting modifications to the electroweak gauge boson properties and investigated the constraints from electroweak precision data, including the $\rho$ parameter and $Z$ pole observables.
Our findings indicate that constraints from $Z$ pole observables are more stringent than those from $\rho$ parameter measurements.

In the \texttt{SMEFT} scenario, constraints from the $\rho$ parameter data require the kinetic mixing parameter to satisfy $\epsilon_6 \lesssim 1.6 \times 10^{-4}$, implying a new physics scale $\Lambda_6 \gtrsim 6.7$ TeV, assuming a dimensionless coefficient $c_6 \sim {\mathcal O}(1)$. On the other hand, constraints from $Z$ pole observables impose a stricter limit of $\epsilon_6 \lesssim 7 \times 10^{-5}$, corresponding to $\Lambda_6 \gtrsim 10$ TeV. Projected precision electroweak data from the CEPC could potentially probe kinetic mixing parameters an order of magnitude smaller than the current bounds.

In the \texttt{SMEFT+$\Sigma$} scenario, kinetic mixing can be significantly enhanced in regions where the triplet VEV is sizable. We found that, if the kinetic mixing parameter is positive, custodial symmetry ({\it i.e.}, $\rho = 1$ at tree level) can be preserved within a finely-tuned parameter space by balancing the kinetic mixing parameter and the triplet VEV. Current $Z$ pole data constrain the kinetic mixing to $-10^{-4} \lesssim \, \epsilon \lesssim 0.001$ and the triplet VEV to $v_\Sigma \lesssim 4$ GeV. Future precision measurements from the CEPC could tighten these bounds by an order of magnitude, probing new physics scales up to approximately $6.5$ TeV ($13$ TeV) assuming a dimensionless coefficient $c_5 = c_6 \sim {\mathcal O}(1)$ ($c_5 = -c_6 \sim {\mathcal O}(1)$).

Before ending, we note the following invariant effective operators in passing:
\bea
\tilde{ {\cal O} }_5 & = & \frac{g g^\prime \tilde c_5 }{\Lambda} \mathrm{Tr} \left[ \Sigma  W_{\mu \nu} \right] \tilde {  B }^{\mu \nu} , \\
\tilde { {\cal O} }_6 & = & \frac{ g g^\prime \tilde c_6 }{\Lambda^2} \mathrm{Tr} \left[ \Phi^{\dagger}  W_{\mu \nu} \Phi \right] \tilde {  B}^{\mu \nu} ,
\eea
where $\tilde{  B}^{\mu\nu}=\frac{1}{2}\varepsilon^{\mu\nu\rho\sigma}  B_{\rho\sigma}$.
After spontaneous symmetry breaking, these operators can produce a mixed term
\[
\frac{\tilde \epsilon }{2}  W^3_{\mu\nu} \tilde{ B}^ {\mu\nu},
\]
with $\tilde{ \epsilon }$ equals $ g g^\prime \tilde c_6 v_\Phi^2 / 2 \Lambda^2$ (\texttt{SMEFT}) or $ - g g^\prime \tilde c_5  v_\Sigma / \Lambda + g g^\prime \tilde c_6 v_\Phi^2 / 2 \Lambda^2$ (\texttt{SMEFT+$\Sigma$}).
This term bears resemblance to the CP-violating $\theta$-term in $SU(2)_L$, making it topological in nature and capable of contributing to the neutron electric dipole moment via non-perturbative effects. The stringent constraints on the $\theta$ parameter from the non-observation of the neutron electric dipole moment also impose tight limits on the parameter $\tilde{\epsilon}$. A detailed investigation of phenomenological implications of $\tilde{\cal O}_5$ and $\tilde{\cal O}_6$ is deferred to future work.

\section*{Acknowledgments}
This work was supported in part by the NSTC grant Nos.  112-2811-M-001-089 (VQT) and 
113-2112-M-001-001 (TCY).

\allowdisplaybreaks
\bibliographystyle{apsrev4-1}
\bibliography{refs}

\end{document}